\documentclass[journal,10pt]{IEEEtran}
\usepackage{algpseudocode}                                 
\usepackage{algorithm}
\usepackage{graphicx}                                      
\usepackage{amsmath}
\usepackage{amssymb}
\usepackage{amsfonts}
\usepackage{booktabs}
\usepackage{multirow}
\usepackage[subrefformat=parens,farskip=0pt,justification=centering]{subfig}
\usepackage{color}
\usepackage{cite}                                          
\usepackage{comment}                                       
\usepackage{soul}                                          
\soulregister\cite7
\soulregister\ref7
\soulregister\pageref7
\usepackage{amsthm}
\usepackage{etoolbox}                                      
\usepackage{url}
\usepackage{nth}                                           
\usepackage{bm}                                            
\usepackage{balance}
\usepackage{threeparttable}
\usepackage[bookmarks=false]{hyperref}                     %
\hypersetup{
    colorlinks = true,
    citecolor  = blue,
    linkcolor  = blue,
    urlcolor   = blue,
}
\usepackage{filecontents}                                  
\usepackage{pgfplots}
\usepackage{pgfplotstable}
\pgfplotsset{compat=newest}
\usepackage[margin=1in]{geometry}                           

\algrenewcommand\textproc{\texttt}

\makeatletter
\let\OldStatex\Statex
\renewcommand{\Statex}[1][3]{%
  \setlength\@tempdima{\algorithmicindent}%
  \OldStatex\hskip\dimexpr#1\@tempdima\relax
}
\makeatother

\RequirePackage[normalem]{ulem} 
\RequirePackage{color}\definecolor{RED}{rgb}{1,0,0}\definecolor{BLUE}{rgb}{0,0,1} 

\newcommand{\tabRef}[1]{TABLE~\ref{#1}}
\newcommand{\figRef}[1]{Fig.~\ref{#1}}

\graphicspath{{../figs/}}

\newcommand\blfootnote[1]{%
  \begingroup
  \renewcommand\thefootnote{}\footnote{#1}%
  \addtocounter{footnote}{-1}%
  \endgroup
}
\begin{document}

\title{
\textbf{
\LARGE{
OpenPARF: An \underline{Open}-Source \underline{P}lacement \underline{a}nd \underline{R}outing Framework for Large-Scale Heterogeneous \underline{F}PGAs with Deep Learning Toolkit
}}
\\
\Large{(Invited Paper)}
}

\author{
    Jing Mai$^{1,2\dagger}$,
    Jiarui Wang$^{1,2\dagger}$,
    Zhixiong Di$^{3}$,
    Guojie Luo$^{1,4}$,
    Yun Liang$^{2,4,5}$,
    Yibo Lin$^{2,4,5*}$ \\
    $^1$School of Computer Science, Peking University \qquad
    $^2$School of Integrated Circuits, Peking University \\
    $^3$School of Information Science and Technology, Southwest Jiaotong University \\
    $^4$Institute of Electronic Design Automation, Peking University, Wuxi, China \\
    $^5$Beijing Advanced Innovation Center for Integrated Circuits, Beijing, China \\
    Email: {\tt \{jingmai,jiaruiwang,gluo,ericlyun,yibolin\}@pku.edu.cn, zxdi@home.swjtu.edu.cn}
}

\maketitle
\thispagestyle{empty} 
\blfootnote{$^\dagger$Equal contribution, ordered by last names. $^*$Corresponding author.}
\begin{abstract}

This paper proposes \texttt{OpenPARF}, an open-source placement and routing framework for large-scale FPGA designs\footnote{
\texttt{OpenPARF} is available at \url{https://github.com/PKU-IDEA/OpenPARF}.
}.
\texttt{OpenPARF} is implemented with the deep learning toolkit \texttt{PyTorch} and supports massive parallelization on GPU.
The framework proposes a novel asymmetric multi-electrostatic field system to solve FPGA placement.
It considers fine-grained routing resources inside configurable logic blocks (CLBs) for FPGA routing
and supports large-scale irregular routing resource graphs.
Experimental results on ISPD 2016 and ISPD 2017 FPGA contest benchmarks and industrial benchmarks demonstrate that \texttt{OpenPARF} can achieve 0.4-12.7\% improvement in routed wirelength and more than $2\times$ speedup in placement.
We believe that \texttt{OpenPARF} can pave the road for developing FPGA physical design engines and stimulate further research on related topics.
\end{abstract}

\section{Introduction}
\label{sec:Introduction}

Computer-Aided Design (CAD) of Field-Programmable Gate Arrays (FPGAs) has been a hot topic in the rapid advancement
and adoption of FPGA technology over the past decades~\cite{chenFPGADesignAutomation2006}.
FPGA CAD flow consists of four major steps: logic synthesis, placement, routing, and bitstream generation~\cite{ghavamiUnravelingIntegrationDeep2023}.
Among these steps, placement and routing (PAR) are two critical steps that affect the overall timing
and power performance of the FPGA design~\cite{chenFPGAPlacementRouting2017}.
It is reported that routing alone accounts for 41-86\% runtime of the entire FPGA CAD flow~\cite{murrayTimingDrivenTitanEnabling2015a}.
Therefore, the quality and efficiency of PAR algorithms highly impact FPGA design closure.

Due to the high heterogeneity of FPGA architectures,
FPGA PAR has several unique characteristics that are different from ASIC.
1) Instance packing.
A collective packing is necessitated in placement to ensconce instances within the designated site.
Packing instances must be subject to sophisticated constraints that vary across FPGA architectures~\cite{PLACE_DAC22_Mai}.
2) Resource heterogeneity.
Sites, categorized by diverse functions such as CLB, DSP, BRAM, IO, etc., exhibit a non-uniform distribution across the FPGA layout~\cite{zhangRapidLayoutFastHard2022}.
3) Large routing scale. The inclusion of fine-grained routing resources with CLBs gives rise to routing
resource graphs with billions of nodes, resulting in substantial routing complexity and runtime overhead~\cite{ROUTE_ASPDAC2023_Wang}.

To achieve high quality and efficiency, the literature has extensively studied PAR algorithms for FPGAs in the past decades\cite{
PLACE_ICCAD2016_Pui_RippleFPGA,
PLACE_TCAD2018_Li,
PLACE_ICCAD2016_Ryan,
PLACE_ICCAD2019_LiLin,
TODAES18_GPlace3_Abuowaimer,
TCAD18_RippleFPGA_Chen,
PLACE_TODAES2018_Li_UTPlaceF2,
PLACE_FPGA2019_Li,
murrayVTRHighperformanceCAD2020,
PLACE_TCAD2020_Chen,
PLACE_ICCAD21_Liang_AMFPlacer,
rajarathnamDREAMPlaceFPGAOpenSourceAnalytical2022,
rajarathnamDREAMPlaceFPGAPLOpenSourceGPUAccelerated2023
}.
However, most of the existing works highly rely on FPGA vendors' CAD tools to obtain indirect feedback
and tightly bind to vendors' architectures,
limiting the flexibility of algorithms and the ability to adapt to new FPGA architectures.

In order to alleviate the burden of reinventing the wheel
and facilitate research on FPGA physical design,
in this paper, we propose \texttt{OpenPARF}, an open-source placement and routing
framework for large-scale heterogeneous FPGAs with deep learning toolkits.
The main contributions are summarized as follows.
\begin{itemize}
\item \texttt{OpenPARF} is an open-source academic FPGA PAR engine that supports complex industrial FPGA architectures with \textit{state-of-the-art} (SOTA) algorithms.
It is implemented with the deep learning toolkit \texttt{PyTorch}, running on both CPU and GPU platforms.
It is highly flexible and extensible for new FPGA architectures and PAR algorithms.
\item  \texttt{OpenPARF} implements the SOTA nonlinear FPGA placement algorithms based on an asymmetrical multi-electrostatic field
system. It is capable of achieving superior placement results under various constraints such as routability, clock feasibility, and SLICEL-SLICEM heterogeneity from advanced FPGA architectures~\cite{PLACE_DAC22_Mai}.
\item  \texttt{OpenPARF} implements a two-stage FPGA routing algorithm. It supports fine-grained CLB-level routing models
and flexible scenarios such as logic pin inequivalence. It is capable of alleviating routing congestion on advanced FPGA architectures effectively~\cite{ROUTE_ASPDAC2023_Wang}.
\end{itemize}
Compared with other SOTA academic PAR engines, \texttt{OpenPARF} can reduce 0.4\%-12.7\% routed wirelength as well as more than $2\times$ speedup in placement efficiency.
We believe that \texttt{OpenPARF} will stimulate the development of FPGA PAR algorithms and foster a surge of research in the future.

\section{The \texttt{OpenPARF} Framework}
\label{sec:Algorithm}
\begin{figure}[tb]
    \centering
    \includegraphics[width=\linewidth]{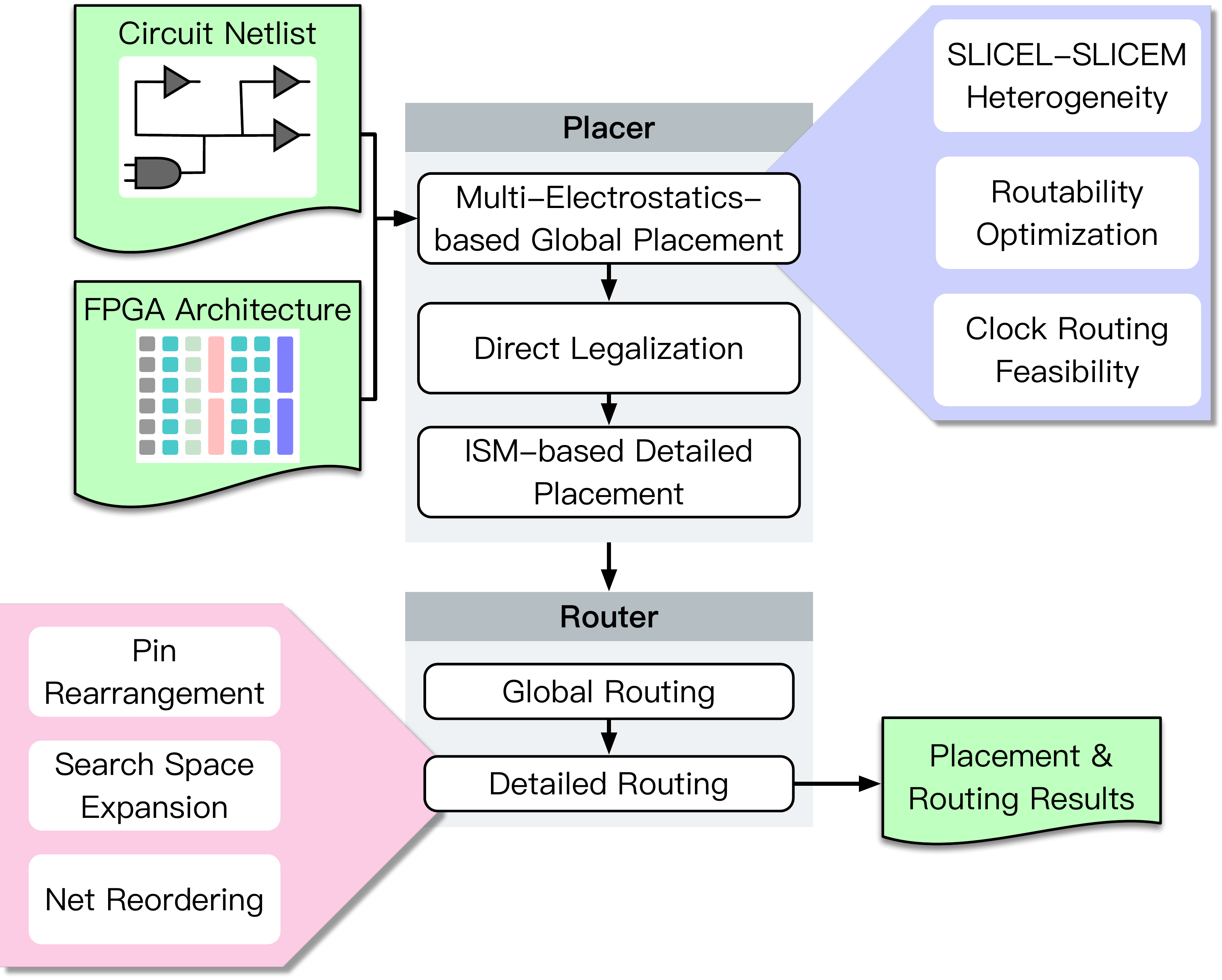}
    \caption{The overall flow of \texttt{OpenPARF}.}
    \label{fig:overall_flow}
 \end{figure}

\subsection{Overall Flow}
The overall flow of \texttt{OpenPARF} is depicted in \figRef{fig:overall_flow}.
\texttt{OpenPARF} consists of two major components: placement and routing.
Firstly, placement information files (bookshelf format) and routing architecture files (XML format like \texttt{VTR} \cite{murrayVTRHighperformanceCAD2020}) are loaded
and constructed into the \texttt{OpenPARF} internal data structure.
Second, based on the internal data structure, \texttt{OpenPARF} employs multi-electrostatic-based FPGA global placement
algorithms and incorporates SLICEL-SLICEM heterogeneity, routability, and clock feasibility into a unified
optimization framework~\cite{PLACE_DAC22_Mai}.
\texttt{OpenPARF} subsequently employs direct legalization algorithm to generate feasible placement results and further improves the placement quality by applying Independent Set Matching (ISM) based detailed placement algorithm~\cite{PLACE_TCAD2019_Li_UTPLACEF_DL}.
Subsequently, the placement results are fed into the two-stage FPGA router in \texttt{OpenPARF}.
Finally, \texttt{OpenPARF} generates feasible routing results through global and detailed routing with three strategies to resolve routing congestion and improve routing quality, i.e., \textit{pin rearrangement}, \textit{search space expansion}, and \textit{net reordering}.
In the following subsections, we will elaborate on the core features of placement and routing.

\subsection{Placement Features}
The objective of placement in \texttt{OpenPARF} is to minimize wirelength while considering SLICEL-SLICEM heterogeneity, routability optimization, and clock routing feasibility.
\subsubsection{SLICEL-SLICEM Heterogeneity}
The CLBs on FPGA can be further categorized into SLICEL and SLICEM.
The logic resources on SLICEL can be configured as LUTs.
Besides LUTs, SLICEM has additional logic resources and can hence be configured as either a distributed RAM or a SHIFT with storage capabilities.
However, if a SLICEM is configured as LUTs, it cannot be utilized as SHIFTs or distributed RAMs, and vice versa.

To support SLICEL-SLICEM heterogeneity, the placer constructs two sets of electric fields: \textit{LUTL} and \textit{LUTM-AL}~\cite{PLACE_DAC22_Mai}.
LUTL models the LUT resources supplied in SLICEL and SLICEM, while LUTM-AL models the additional logic resources supplied in SLICEM, but not in SLICEL.
A LUT instance only utilizes resources in the LUTL field, whereas a distributed RAM or SHIFT instance utilizes resources in both the LUTL and LUTM-AL fields.
This field setup prevents a distributed RAM or SHIFT instance from being placed in SLICEL sites,
but allows a LUT instance to be placed in both SLICEL and SLICEM sites.

\subsubsection{Routability Optimization}
Routability reveals whether the current placement is difficult or impossible to route given available target device routing resources.
Our placer adopts the area inflation-based routability optimization strategy~\cite{PLACE_TCAD2021_Meng} to improve the routability of placement results.
Instances in congested regions are inflated by a greater factor.
Therefore, the placer can effectively reduce the routing congestion and improve the routability of the placement result.

\subsubsection{Clock Routing Feasibility}
Advanced FPGAs have dedicated clock routing resources to route clock signals.
Take Xilinx \textit{UltraScale VU095} as an example of FPGA devices.
The target FPGA device is divided into rectangular-shaped clock regions (CRs) in a grid manner, and each CR can be further subdivided into pairs of lower and upper half columns (HCs).
The clock architecture imposes clock routing constraints on placement, the clock region constraint, and the half-column constraint~\cite{BENCH_ISPD2017_PLACE}.
Our placer adopts the clock network planning algorithm in~\cite{PLACE_DAC22_Mai}, and the clock feasibility is incorporated into the placement objective function as a penalty term.

\subsubsection*{Nested Lagrangian Relaxation}
\texttt{OpenPARF} adopts the nested Lagrangian relaxation algorithm~\cite{PLACE_DAC22_Mai} to solve the placement problem with SLICEL-SLICEM heterogeneity, routability optimization, and clock routing feasibility.
SLICEL-SLICEM heterogeneity is modeled by the \textit{LUTL} and \textit{LUTM-AL} electric field systems, which are incorporated into the electrostatic density function.
Routing congestion is gradually resolved by inflating the area of instances in congested regions.
The clock routing feasibility is ensured by the clock network planning algorithm.
The placement problem is solved by iteratively solving the relaxed placement problem and updating the Lagrangian multipliers until the placement result is feasible.

\begin{table*}[tb]
  \centering
  \caption{Placement runtime (PRT in seconds), routing runtime (RRT in minutes), and routed wirelength ($\times10^4$ RWL) comparison on ISPD 2016 benchmarks~\cite{BENCH_ISPD2016_PLACE}.}
    \begin{tabular}{|c|c|ccc|ccc|ccc|}
    \hline
      \multirow{2}[2]{*}{Design} & \multirow{2}[2]{*}{\#LUT/\#FF/\#RAM/\#DSP} & \multicolumn{3}{c|}{\texttt{RippleFPGA}~\cite{TCAD18_RippleFPGA_Chen}} & \multicolumn{3}{c|}{\texttt{DREAMPlaceFPGA}~\cite{rajarathnamDREAMPlaceFPGAPLOpenSourceGPUAccelerated2023}} & \multicolumn{3}{c|}{\texttt{OpenPARF}} \\
          &       & PRT   & RRT   & RWL   & PRT   & RRT   & RWL   & PRT   & RRT   & RWL \\
    \hline
    \hline
      \texttt{FPGA01} & 50K/55K/0/0 & 41    & 3     & 36.44 & 32    & 3     & 31.78 & 39    & 3     & 31.72 \\
      \texttt{FPGA02} & 100K/66K/100/100 & 64    & 5     & 75.29 & 57    & 5     & 68.17 & 59    & 5     & 67.73 \\
      \texttt{FPGA03} & 250K/170K/600/500 & 245   & 17    & 346.91 & 108   & 15    & 299.56 & 120   & 15    & 294.75 \\
      \texttt{FPGA04} & 250K/172K/600/500 & 337   & 22    & 632.96 & 97    & 22    & 569.85 & 112   & 22    & 577.30 \\
      \texttt{FPGA05} & 250K/174K/600/500 & 391   & 57    & 1222.46 & 91    & 56    & 1167.60 & 123   & 54    & 1148.72 \\
      \texttt{FPGA06} & 350K/352K/1000/600 & 593   & 25    & 652.41 & 183   & 29    & 571.11 & 218   & 27    & 573.54 \\
      \texttt{FPGA07} & 350K/355K/1000/600 & 782   & 46    & 1106.96 & 159   & 51    & 964.44 & 209   & 45    & 965.24 \\
      \texttt{FPGA08} & 500K/216K/600/500 & 490   & 40    & 958.26 & 146   & 36    & 911.67 & 184   & 38    & 896.91 \\
      \texttt{FPGA09} & 500K/366K/1000/600 & 738   & 58    & 1327.34 & 191   & 54    & 1203.49 & 260   & 50    & 1198.22 \\
      \texttt{FPGA10} & 350K/600K/1000/600 & 1180  & 28    & 711.48 & 180   & 28    & 544.52 & 258   & 35    & 542.02 \\
      \texttt{FPGA11} & 480K/363K/1000/400 & 721   & 58    & 1281.65 & 148   & 56    & 1250.49 & 221   & 59    & 1253.78 \\
      \texttt{FPGA12} & 500K/602K/600/500 & 883   & 40    & 761.37 & 184   & 37    & 674.21 & 290   & 38    & 670.94 \\
    \hline
    \hline
    Ratio &       & 2.771 & 1.015 & 1.127 & 0.786 & 1.000 & 1.004 & 1.000 & 1.000 & 1.000 \\
    \hline
    \end{tabular}%
  \label{tab:ispd2016_results}
\end{table*}

\subsection{Routing Features}
The router of our \texttt{OpenPARF} framework consists of two stages, coarse-grained level global routing and fine-grained level detailed routing~\cite{ROUTE_ASPDAC2023_Wang}. In the global routing phase, we generate an inter-site level routing result to guide the behavior of the detailed router. In the detailed routing phase, we generate the logic element routing result at both inter-site and intra-site levels.

\subsubsection{Global Routing}

The target of the global router is to generate inter-site level coarse-grained routing results to guide the follow-up detailed routing.
global router follows the \texttt{Pathfinder} algorithm by abstracting the FPGA layout as a grid graph.
logic site is represented as a vertex, and the routing channels are represented as edges in the grid graph.
Two vertices are connected if there are routing channels connecting them.
The capacity of an edge is the width of the routing channel between its two logic sites, and the edge capacity is set to the routing channel width.

\subsubsection{Detailed Routing}
The target of the detailed router is to generate the inter- and intra-site routing paths under the guidance of the global routing results.
Detailed router follows a similar negotiation-based algorithm to the global router, but abstracts the FPGA layout to a more fine-grained routing resource graph (RRG).
In the detailed router's RRG, the vertex represents the logic pin inside the FPGA, and the directed edge represents the logic connection between two logic pins.

Detailed router applies three strategies to efficiently and effectively generate detailed routing results.
1) To deal with routing congestion at the input of each LUT, the detailed router applies \textit{pin rearrangement} technique.
Detailed router regards the input logic pins of a LUT as a vertex with large capacity, and rewrites the truth table of the LUT after detailed routing.
2) Detailed router dynamically applies \textit{search space expansion} technique to restrict the search space.
3) To effectively resolve the routing congestion between different logic nets, the detailed router adopts the \textit{net reordering} strategy at the beginning of each rip-up and reroute iteration.



\section{Experimental Results}
\label{sec:Results}

 \begin{table*}[tb]
    \centering
    \caption{Placement runtime (PRT in seconds), routing runtime (RRT in minutes), and routed wirelength ($\times10^4$ RWL) comparison on ISPD 2017 benchmarks~\cite{BENCH_ISPD2017_PLACE}.}
        \begin{tabular}{|c|cc|ccc|ccc|}
        \hline
        \multirow{2}[2]{*}{Design} & \multirow{2}[2]{*}{\#LUT/\#FF/\#BRAM/\#DSP} & \multirow{2}[2]{*}{\#Clock} & \multicolumn{3}{c|}{\texttt{RippleFPGA}~\cite{TCAD18_RippleFPGA_Chen}} & \multicolumn{3}{c|}{\texttt{OpenPARF}} \\
              &       &       & PRT   & RRT   & RWL   & PRT   & RRT   & RWL \\
        \hline \hline
        \texttt{CLK-FPGA01} & 211K/324K/164/75 & 32    & 278   & 10    & 238.54 & 131   & 10    & 205.44 \\
        \texttt{CLK-FPGA02} & 230K/280K/236/112 & 35    & 250   & 15    & 261.85 & 127   & 14    & 246.65 \\
        \texttt{CLK-FPGA03} & 410K/481K/850/395 & 57    & 537   & 24    & 648.69 & 206   & 24    & 594.00 \\
        \texttt{CLK-FPGA04} & 309K/372K/467/224 & 44    & 346   & 18    & 440.09 & 157   & 19    & 420.30 \\
        \texttt{CLK-FPGA05} & 393K/469K/798/150 & 56    & 501   & 25    & 560.18 & 201   & 23    & 510.62 \\
        \texttt{CLK-FPGA06} & 425K/511K/872/420 & 58    & 545   & 28    & 678.43 & 218   & 28    & 617.28 \\
        \texttt{CLK-FPGA07} & 254K/309K/313/149 & 38    & 288   & 13    & 276.29 & 136   & 13    & 256.62 \\
        \texttt{CLK-FPGA08} & 212K/257K/161/75 & 32    & 235   & 10    & 213.06 & 119   & 10    & 196.67 \\
        \texttt{CLK-FPGA09} & 231K/358K/236/112 & 35    & 312   & 13    & 297.02 & 148   & 14    & 250.98 \\
        \texttt{CLK-FPGA10} & 327K/506K/542/255 & 47    & 465   & 23    & 544.07 & 195   & 14    & 451.28 \\
        \texttt{CLK-FPGA11} & 300K/468K/454/224 & 44    & 421   & 24    & 516.67 & 182   & 30    & 421.52 \\
        \texttt{CLK-FPGA12} & 277K/430K/389/187 & 41    & 378   & 18    & 403.59 & 167   & 20    & 336.03 \\
        \texttt{CLK-FPGA13} & 339K/405K/570/262 & 47    & 393   & 21    & 464.78 & 178   & 19    & 428.41 \\
        \hline \hline
        Ratio &       &       & 2.251 & 1.037 & 1.128 & 1.000 & 1.000 & 1.000 \\
        \hline
        \end{tabular}%
    \label{tab:ispd2017_results}
  \end{table*}

\texttt{OpenPARF} is implmented in C++ and Python along with the open-source machine learning toolkit \texttt{Pytorch} for fast gradient computation~\cite{PLACE_DAC2019_Lin}.
We conduct experiments on a Linux server that consists of an Intel(R) Xeon(R) Gold 6230 CPU @ 2.10GHz (40 cores), one NVIDIA RTX 2080Ti GPU, and 512GB memory.
We demonstrate the effectiveness and efficiency of our proposed algorithm on both ISPD 2016~\cite{BENCH_ISPD2016_PLACE} and ISPD 2017~\cite{BENCH_ISPD2017_PLACE} academic benchmarks and industrial benchmarks.
The industrial benchmarks are provided by our industrial collaborator.
We also complete the routing architecture of the academic benchmarks under their guidance.
\tabRef{tab:ispd2016_results}, \tabRef{tab:ispd2017_results} and \tabRef{tab:industry_results} provide the overview of evaluation benchmarks.
The benchmarks encompass various instance types and netlist sizes ranging from 21K to 1100K.

\subsection{Evaluation on Academic Benchmarks}

\tabRef{tab:ispd2016_results} and \tabRef{tab:ispd2017_results} elucidate the comparative results between \texttt{OpenPARF} and two SOTA FPGA placers, \texttt{RippleFPGA}~\cite{TCAD18_RippleFPGA_Chen} and \texttt{DREAMPlaceFPGA}~\cite{rajarathnamDREAMPlaceFPGAPLOpenSourceGPUAccelerated2023}
\footnote{
\texttt{DREAMPlaceFPGA} includes two works, namely \cite{rajarathnamDREAMPlaceFPGAOpenSourceAnalytical2022} and \cite{rajarathnamDREAMPlaceFPGAPLOpenSourceGPUAccelerated2023}.
In this paper, the placement runtime of \texttt{DREAMPlaceFPGA} is extracted from the latest work \cite{rajarathnamDREAMPlaceFPGAPLOpenSourceGPUAccelerated2023}.
}.
It is noteworthy that \texttt{RippleFPGA} and \texttt{DREAMPlaceFPGA} embody the epitome of analytical placement algorithms,
one based on quadratic programming,
and the other on non-convex optimization models.
By incorporating the placement results of the aforementioned placers into \texttt{OpenPARF}'s router,
We proceed to access placement runtime (PRT), routing runtime (RRT), and routed wirelength (RWL) between them, with routing time serving as an indicator of routability.

\subsubsection{Evaluation on ISPD 2016 Benchmarks}

The experimental results on the ISPD 2016 experiment unveil \texttt{OpenPARF}'s advantages over other placers, showcasing a reduction in wirelength by 12.7\% and 0.4\%, respectively, coupled with up to $2.77\times$ placement speedup.
Notably, \texttt{DREAMPlaceFPGA}  harnesses GPU acceleration for legalization.
\texttt{OpenPARF} exhibits a 21.4\% decrease in runtime when compared with \texttt{DREAMPlaceFPGA}.
We believe that \texttt{OpenPARF} has vast potential for further performance enhancement by delving into the realms of heterogeneous parallelization, accelerating both legalization and detailed placement.

\subsubsection{Evaluation on ISPD 2017 Benchmarks}

We further conducted a comparative analysis between \texttt{RippleFPGA} and \texttt{OpenPARF} on ISPD2017 benchmarks.
As \texttt{DREAMPlaceFPGA} does not support clock routing constraints in ISPD2017 benchmarks, we do not include it in the comparison.
The experimental results demonstrate that \texttt{OpenPARF} outperforms \texttt{RippleFPGA} by achieving a 12.8\% reduction in wirelength and with $2.251\times$ speedup.
It is noteworthy that \texttt{OpenPARF} exhibits a reduction of 3.7\% in routing runtime compared to \texttt{RippleFPGA}, indicating that \texttt{OpenPARF} is capable of generating routing results with favorable routability even under intricate constraints.


\subsection{Evaluation on Industrial Benchmarks}

\tabRef{tab:industry_results} presents the placement runtime, routing runtime, and routed wirelength of \texttt{OpenPARF} on industrial benchmarks, which encompass distributed RAMs and SHIFTs, exhibiting the heterogeneous nature of SLICEL-SLICEM constraints.
The experimental results show the performance and efficiency of \texttt{OpenPARF} when dealing with SLICEL-SLICEM constraints.
\texttt{OpenPARF} has demonstrated commendable efficacy by leveraging an asymmetric multi-electrostatic system to address the SLICEL-SLICEM heterogeneity.

\section{Conclusion and Future Work}
\label{sec:Conclusion}
\begin{table}[tb]
  \centering
  \caption{Placement runtime (PRT in seconds), routing runtime (RRT in minutes), and routed wirelength ($\times10^3$ RWL) on industrial benchmarks.}
  \resizebox*{\linewidth}{!}{
\begin{tabular}{|c|ccc|ccc|}
\hline
\multirow{2}{*}{Design}         & \multirow{2}{*}{\begin{tabular}[c]{@{}c@{}}\#LUT/\#FF/\\ \#BRAM/\#DSP\end{tabular}} & \multirow{2}{*}{\begin{tabular}[c]{@{}c@{}}\#Distributed\\ RAM + \#Shift\end{tabular}} & \multirow{2}{*}{\#Net} & \multicolumn{3}{c|}{\texttt{OpenPARF}} \\ \cline{5-7} 
                                &                                                                                     &                                                                                      &                        & PRT                & RRT             & RWL              \\ \hline \hline
\texttt{IND01} & 17K/11K/0/13                                                                        & 9                                                                                    & 52492                  & 72.36              & 10              & 90               \\
\texttt{IND02} & 11K/10K/0/24                                                                        & 6                                                                                    & 26678                  & 77.82              & 15              & 100              \\
\texttt{IND03} & 109K/12K/0/0                                                                        & 0                                                                                    & 121554                 & 109.54             & 108             & 1021             \\
\texttt{IND04} & 29K/17K/0/16                                                                        & 218                                                                                  & 60968                  & 69.39              & 19              & 283              \\
\texttt{IND05} & 64K/191K/64/928                                                                     & 29K                                                                                  & 371808                 & 126.38             & 109             & 2360             \\
\texttt{IND06} & 112K/65K/21/0                                                                       & 0                                                                                    & 221182                 & 88.28              & 176             & 1593             \\
\texttt{IND07} & 40K/156K/89/768                                                                     & 26K                                                                                  & 294075                 & 140.33             & 68              & 1450             \\ \hline
\end{tabular}
  }
  \label{tab:industry_results}
\end{table}

This paper introduces \texttt{OpenPARF}, an open-source placement and routing framework for large-scale FPGAs.
Built upon the deep learning toolkit \texttt{PyTorch}, \texttt{OpenPARF} supports GPU acceleration with agile and flexible programming interfaces.
In placement, \texttt{OpenPARF} embodies the SOTA asymmetrical multi-electrostatic FPGA placement algorithms and harnesses the nested Lagrangian relaxation methodology to resolve SLICEL-SLICEM heterogeneity, routability and clock routing feasibility.
In routing, \texttt{OpenPARF} implements a two-stage FPGA routing algorithm, which supports fine-grained CLB-level routing models and flexible scenarios like logic pin inequivalence.
We have a belief that \texttt{OpenPARF} can serve as a crucial platform for exploring next-generation high-performance FPGA placement and routing algorithms, and stimulates the development of future FPGA CAD algorithms.

\section*{Acknowledgement}
This work was supported in part by the National Science Foundation of China (Grant No. T2293700 and T2293701) and the 111 Project (B18001).

{
\bibliographystyle{IEEEtran}
\bibliography{./ref/merged}
}

\end{document}